\documentstyle[twoside,fleqn,espcrc2]{article}

\newcommand{\AmS}{{\protect\the\textfont2
  A\kern-.1667em\lower.5ex\hbox{M}\kern-.125emS}}

\title{APEmille: a parallel processor in the teraflop range}

\author{E. Panizzi for the APE collaboration\\
Gruppo APE, Istituto Nazionale di Fisica Nucleare, P.le
A. Moro,2 I-00185 Roma, Italy\\
{\tt panizzi@roma1.infn.it\\
http://chimera.roma1.infn.it/ape.html
}}

\begin{document}

\begin{abstract}
APEmille is a SIMD parallel processor under development at the
Italian National Institute for Nuclear Physics (INFN). It is the third
machine of the APE family, following Ape and Ape100 and delivering peak 
performance in the Tflops range.
APEmille is very well suited for Lattice QCD applications, both for
its hardware characteristics and for its software and language
features.
APEmille is an array of custom arithmetic processors arranged on 
a tridimensional torus. The replicated processor is a pipelined VLIW
device performing integer and single/double precision IEEE floating
point operations. The processor is optimized for complex computations and
has a peak performance of 528Mflop at 66MHz. Each replica has
8 Mbytes of locally addressable RAM.
In principle an array of 2048 nodes is able to break the Tflops barrier.
Two other custom processors are used for program flow control, global
addressing and inter node communications. Fast nearest neighbour
communications as well as longer distance communications and data
broadcast are available.
APEmille is interfaced to the external world by a PCI interface and a
HIPPI channel. A network of PCs act as the host computer. The APE operating
system and the cross compiler run on it.
A powerful programming language named TAO is provided and is highly
optimized for QCD. A C++ compiler is foreseen.
The TAO language is as simple as Fortran but as powerful as object
oriented languages. Specific data structures, operators and even
statements can be defined by the user for each different application.
Effort has been made to define the language constructs for QCD.
\end{abstract}

\maketitle

\section{INTRODUCTION}

APEmille is a parallel processor oriented to the lattice calculation
and suitable for massively parallel homogeneous problems.
It is a SIMD array processor with local addressing capability and has
a tridimensional toroidal topology with efficient communications.

APEmille \cite{proposal,addendum} is the third computer of the APE
family, currently under development. The previous ones - Ape
\cite{Aperef} and Ape100 \cite{Ape100H,Ape100S} - had similar
characteristics, although they did not support many APEmille
capabilites, first of which the local addressing and the hardware
double precision, and although they were much less powerful: in fact
APEmille peak power is in the teraflop range, while Ape100 was a
100Gflop machine and Ape was a 1 Gflop one.

The main target of this machine is the quantum chromo dynamics
simulations, and big effort has been made to optimize the performance
in that area and to achieve easy programmability by
tuning the programming language with some QCD oriented extensions.
Nevertheless many other applications are good candidates both for the
machine architecture and the programing environment: fluid dynamics,
seismic migration, atmosphere, neural networks.

\section{ARCHITECTURAL STRUCTURE}

APEmille is based on a number of architectural keypoints:
\begin{itemize}
\item
it is a Single Instruction Multiple Data (SIMD) machine, in fact
each node performs the same instruction at the same time on its own
data;
\item
it is based on Very Long Instruction Word (VLIW) processors, in which
each field of the instruction word drives a different internal device
of the processor, thus exploiting internal parallelism;
\item
it has a Harvard architecture, in fact program memory and data
memories are physically different; the data memory is distributed,
each node has its own one;
\item
it is based on the replication of three custom processors, named
Tarzan, Jane and Cheetah: Tarzan, which is replicated every eight
nodes, drives the program flow and issues the global addressing to its
nodes; Jane is the node processor and is capable of floating
point as well as integer computations; Cheetah is the
communication processor and is replicated every eight nodes as well.
\end{itemize}

Tarzan drives the program flow for all the nodes taking into
account {\em global conditions} (results of its own comuptations) as well as
aggregate evaluation of {\em local conditions} (those computed by the
Janes). Tarzan computes global addresses and delivers them to the
nodes.

Jane is capable of performing the basic operation $a\times b+c $
called {\tt normal} which can have either real or complex operands.
In fact Jane can manage real and complex numbers in hardware.
Real numbers can be both in single and double precision, while the
hardware representation of complex numbers is single precision.
Jane is capable of starting a new normal operation at each clock cycle
and has an eight stage pipeline for double precision numbers and a six
stage one for single precision.
Integer, logical and bitwise operations are performed in shorter
pipes.
Jane has local addressing capability: in fact every node can add a
different local offset to the global address provided by Tarzan.

Cheetah can perform first neighbour and longer distance homogeneus
communications (in which the distance between senders and receivers is
the same for all nodes), broadcast communications (in which only a few
nodes are senders, while all the others are receivers), and - not in
the first APEmille release - unhomogeneus communications, in which
each node can send (or ask for) data to (from) a node at an arbitrary
distance. In the last kind of communications conflicts on the
destinations may occur, so this communication model will require some
additional hardware in the Cheetah processors.

APEmille is connected to the external world via the APEChannel, a
32-bit channel that connects Tarzan to a PCI card or a HIPPI
driver. Thus any PCI based computer can be used as host for APEmille
to upload programs, upload/download data and run all the software
tools. Either the host or a specialized device (such as a disk array
accessible via HIPPI) can be used for mass storage.

\section{HARDWARE CHARACTERISTICS}

APEmille is scalable from eight processing nodes up to 2048. In fact
eight nodes are assembled on the same board together with their
communication and driving logic, so only multiples of eight nodes are
possible. The eight nodes are connected as if they were on the vertices of a
cube, in a 2$\times$2$\times$2 configuration. Intermediate size machines will have
2$\times$2$\times$8, 2$\times$8$\times$8 and 8$\times$8$\times$8 configurations. The full APEmille configuration
is 32$\times$8$\times$8. The APEmille custom processors are designed by the APE Group
and implemented as standard cell ASIC design metodology.

Tarzan is only capable of integer computations and has a dedicated
device for address computation, the AGU (Address Generation Unit).
Tarzan has a register file (multiport internal RAM) with 64 register (32
bits each). It has its own static data memory and drives the
progam memory, which also feeds Jane with microcode. SDRAM (synchronous
dynamic RAM) technology is used for program memory.
 
Jane is capable of integer and floating point computations. It has a
512-deep register file with five ports: three output ports (used for
the three operands in a {\tt normal} operation), one input port (to
store the result) and one input/output port (to excange data with
memory or with the other nodes). At each clock cycle Jane is capable
of starting one {\tt normal} operation and exchanging one 32 bit word
with the external world. The memories used for Jane data are SDRAM.

Cheetah is a bitsliced device: four identical chips are required on
each board to drive the communication of the 32bit words exchanged
among the Janes. Each chip manages one quarter of the word.

\begin{table*}[hbt]
% space before first and after last column: 1.5pc
% space between columns: 3.0pc (twice the above)
\setlength{\tabcolsep}{1.5pc}
% -----------------------------------------------------
% adapted from TeX book, p. 241
\newlength{\digitwidth} \settowidth{\digitwidth}{\rm 0}
\catcode`?=\active \def?{\kern\digitwidth}
% -----------------------------------------------------
\caption{APEmille peak performances and memory sizes}
\label{tab:perfmem}
\begin{tabular*}{\textwidth}{@{}l@{\extracolsep{\fill}}rrrr}
\hline
                 & \multicolumn{2}{l}{Peak performances (flop/s)} 
                 & \multicolumn{2}{l}{Data memory size (bytes)} \\
\cline{2-3} \cline{4-5}
                 & \multicolumn{1}{r}{66MHz clock cycle} 
                 & \multicolumn{1}{r}{100MHz clock cycle} 
                 & \multicolumn{1}{r}{Min} 
                 & \multicolumn{1}{r}{Max}         \\
\hline
Node     & $  528M$ & $  800M$ & $   8M$ & $ 32M$ \\
Board    & $4.224G$ & $  6.4G$ & $  64M$ & $256M$ \\
Subcrate & $ 16.9G$ & $ 25.6G$ & $ 256M$ & $  1G$ \\
Crate    & $ 67.5G$ & $102.4G$ & $   1G$ & $  4G$ \\
Tower    & $  270G$ & $409.6G$ & $   4G$ & $ 16G$ \\
APEmille & $1.081T$ & $  1.6T$ & $ 1 6G$ & $ 64G$ \\
\hline
\end{tabular*}
\end{table*}

\section{APEmille NUMBERS}

APEmille clock frequency is 66MHz in the first release and will be
100MHz in the final release. As Jane is capable of starting a new {\tt
normal} operation each clock cycle, it performs 66M up to 100M normal
operations per second. When dealing with complex numbers, a normal
operation consists in 8 floating point operations, so Jane's peak
power is 528Mflop/s in the first release and 800Mflop/s in the final
one. These numbers are especially valid for QCD.  Performances for the
different configurations are described in table \ref{tab:perfmem}. The
nodes' memory sizes of the different configurations are described in the
table as well. The program memory is 512KWords 176 bits each
(due to the VLIW architecture). Tarzan data memory is a 256Kbytes static
RAM. The APEChannel will be driven by a 33MHz clock, thus providing a
bandwidth of 133MB/s.

\section{SOFTWARE AND LANGUAGES}

Three kinds of software tools will be provided with APEmille. The
first are the language related tools: cross compilers, optimisers, 
libraries. The second ones are the Operating System related tools,
i.e. the OS itself, the monitor program, and the graphic symbolic
debugger. The last group is composed of the simulator and the graphic
profiler.
Part of this software is already working: the compilation chain, from
the assembler downto the machine dependent optimizer, is already
producing executable code that runs on the simulator in a single board
configuration. Part of the operating system is supporting the
simulator work.
The machine independent optimizer is being developed and will be the
core of the APEmille high level language compilers.
Two compilers are foreseen, the TAO compiler and a C++ compiler.
TAO is the APE extensible language with fortran flavour. This language
allows the user to define new operators and new statements or overload old
ones.
This means that it is possible to extend the TAO language toward an
application oriented language so that codes become shorter, more comprehensible
and need less comments and documentation. Moreover maintenance and
upgrade of application codes becomes easier.
An example of extension, on which we spent much effort, is the QCD
header file. This file, included by QCD application programs, extends
the TAO language with data structures suitable for the QCD (like
spinor, su3 etc.) and operators applicable to these data types.

\end{document}